# Multi-Frame Energy-Selective Imaging System for Fast-Neutron Radiography

Volker Dangendorf, Doron Bar, Benjamin Bromberger, Gennady Feldman, Mark B. Goldberg, Ronald Lauck, Ilan Mor, Kai Tittelmeier, David Vartsky, Mathias Weierganz

*Abstract*—A new instrument for high resolution imaging of fast-neutrons is presented here. It is designed for energy selective radiography in nanosecond-pulsed broad-energy (1 - 10 MeV) neutron beams. The device presented here is based on hydrogenous scintillator screens and single- or multiple-gated intensified camera systems (ICCD). A key element is a newly developed optical amplifier which generates sufficient light for the high-speed intensified camera system, even from such faint light sources as fast plastic and liquid scintillators. Utilizing the Time-of-Flight (TOF) method, the detector incorporating the above components is capable of simultaneously taking up to 8 images, each at a different neutron energy.

## I. INTRODUCTION

Most established X- and γ-ray methods for explosives detection depend critically on shape recognition and therefore on human operator skill. Furthermore, photon radiography only permits very limited differentiation among elements in the low-Z range and the latter can be rendered quasi undetectable by high-Z element shielding. In contrast, Fast-Neutron Resonance Radiography (FNRR) is one of the most promising methods for fully automatic detection and identification of explosives concealed in luggage and cargo. Among others, this is related to the fact that neutron transmission depends only weakly on absorber Z and neutrons readily penetrate high-Z materials. Furthermore, neutrons probe the nuclear properties of the absorber and exhibit highly characteristic structure in their various interaction cross sections with different isotopes and at diverse neutron energies. Indeed, in the 'eighties and 'nineties, these features were exploited in developing various interrogation techniques for automatic luggage and cargo inspection, using e.g. neutron-induced, element-specific gamma emission (TNA with thermal neutrons. PFNA with fast neutrons), or energy-dependent characteristic neutron transmission (PFNTS).

This work was supported in part by the U.S. Transportation Security Laboratory (TSL) under Grant HSTS04-05-R-RED108.
V. Dangendorf is affiliated to Physikalisch-Technische Bundesanstalt, Braunschweig, Germany (phone: +49-531-5926510; fax: +49-531-5926405; e-mail: volker.dangendorf@ptb.de), as are B. Bromberger, R. Lauck, K. Tittelmeier and M. Weierganz
D. Bar, G. Feldman, M. B. Goldberg and D. Vartsky are affiliated to Soreq NRC, Yavne, Israel

Recent reviews of various neutron-based inspection methods can be found in [1-3].

In this work we focus on FNRR, which exploits the isotope-specific energy dependence of the total neutron cross sections to measure elemental distributions and identify hazardous materials in luggage and cargo. For example, Fig. 1 shows the calculated transmission spectra of MeV neutrons through 10 cm material of water, polyethylene and Tri-Acetone Tri-Peroxide (TATP). TATP is an improvised explosive made of standard household chemicals, that has been employed by terrorists in Israel, the U.K. and a recently foiled attempt in Germany. The various dips and bumps in each spectrum are characteristic of its elemental composition, and specifically of its relative abundance of carbon, nitrogen and oxygen.

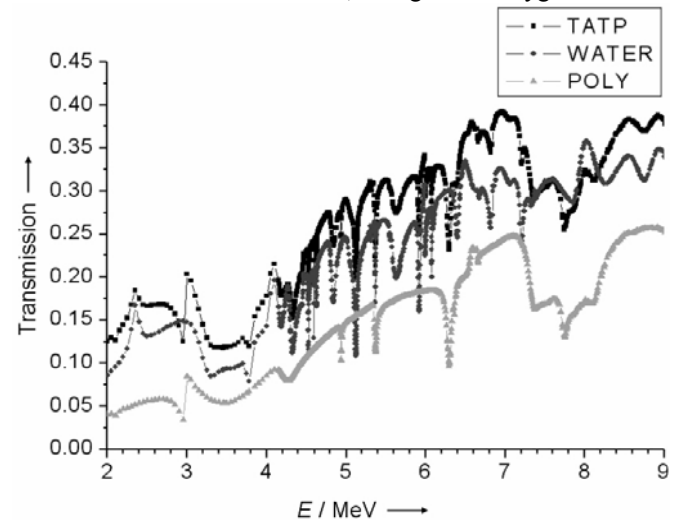

Fig. 1. Calculated transmission through 10 cm thick TATP, water and polyethylene

A pre-requisite for FNRR is precise knowledge of neutron energy. Two approaches were proposed to achieve this goal:
a) sequentially generating transmission images at various quasi-monoenergetic neutron energies [4] or b) using a broad energy pulsed neutron beam and Time-of-Flight methods for energy selection. The latter, also known as Pulsed Fast-Neutron Transmission Spectroscopy (PFNTS) was initiated by the Oregon group in 1985 [5 - 7] and further developed by Tensor Technology, Inc [8 - 13]. It is based on a pulsed broad energy neutron beam, usually produced by an accelerator-based point-like source, as well as a large-area neutron imaging system with Time-of-Flight (TOF) capability. With these, multiple neutron transmission images of inspected items can be acquired essentially simultaneously.

In the late 1990s, an advanced PFNTS system built by Tensor was taken through several blind tests for the U.S. Federal Aviation Administration (FAA) [8-13]. The results of the blind tests indicated that PFNTS is a powerful technique and provides a robust detection method for bulk explosives, that could quite conceivably exceed the performance of next-generation CT-based systems.

Despite its high potential, a U.S. National Materials Advisory Board Panel of the National Research Council recommended in 1999 [13] not to deploy the PFNTS system at airports, since the relatively poor spatial resolution achievable at that time (~4 x 4 cm$^2$) was inadequate for reliable detection of thin sheet explosives. Furthermore, false alarm rates were relatively high. Finally, size, weight and safety aspects were an issue.

Nevertheless, the panel did explicitly stress that its conclusions were primarily dictated by the lack of supporting technologies (detectors with better spatial resolution and more suitable neutron sources), rather than by any inherent shortcomings of the PFNTS method itself. The panel concluded that a mature system of this type, being completely automatic, would represent a significant improvement over current systems.

As a result, PFNTS activity subsequently focused on the development of components, such as a more compact neutron source, jointly being developed by Tensor and Lawrence Berkeley National Laboratory (LBNL) [15]. However, Tensor's PFNTS detectors are still incapable of proffering the spatial resolution required for reliable detection of thin explosives.

In this paper we present the concept and first results of our work on a high resolution fast-neutron imaging system with TOF-spectrometry capability. In recent years we have investigated several approaches, including gas-detectors and scintillator-based optical detectors operating in neutron counting mode [16,17].

Specifically, we describe the concept of the Time-Resolved Integrative Optical Neutron detector (TRION), which allows for energy-selective neutron imaging at very high neutron fluxes (hitherto inaccessible by means of event-counting methods) and is based primarily on commercially available electro-optical components. A comprehensive overview of our work on novel detectors for this purpose can be found in [16-20] and a detailed recent publication on TRION is soon to be published in JINST [21].

## II. Detector Concept

### A. Detector requirements

A practicable detector for an aviation-cargo container or palletized luggage inspection system needs to fulfill several requirements, in order to enable sufficiently short inspection times (< 5 min) and to take images from several projection angles. The latter are required to obtain spatial information on the objects, in particular also for detecting thin sheet explosives. The overall size of the detector should be about 20 x 150 cm$^2$ – so that the cargo item can be efficiently scanned. Detection efficiency should be higher than 10 % and in fact, as high as possible, without giving rise to excessive build-up of scattered neutrons inside the neutron converter. The detector should be capable of handling neutron rates in excess of $10^6 s^{-1} cm^{-2}$. PFNTS requires energy resolution of several hundred keV (FWHM) in the neutron energy range 1 – 10 MeV (and specifically, ~500 keV at 8 MeV). Energy information is provided by measuring neutron TOF. Therefore the system requires a time resolution which depends on the flight distance from the neutron source to the detector. At a TOF distance of 5 m this demands ~4 ns FWHM time resolution. To also ensure good detection capability for sheet explosives, spatial resolution of 1–2 mm (FWHM) is the design goal.

### B. Concept of the single frame TRION Detector

The basic design of the TRION detector is shown in Fig. 2. Since this first prototype is capable of imaging only a single TOF-frame at a time, it will be referred to here as Generation 1 (Gen1) TRION. In this version neutrons arrive at the detector after their TOF, interact in a scintillating slab or a fiber array screen (fiber diameters of 0.25 and 0.5 mm were available), producing an image in which the spatial neutron distribution is converted to a corresponding visible light image. This light image is viewed through a high aperture lens by a very fast gateable image intensifier. The intensifier not only amplifies the light intensity but, more importantly, acts as an electronic shutter that is opened for a gate period of $\Delta t$ as short as 4 – 10 ns at a pre-selected TOF relative to the beam burst. Repetition rates for the beam pulses were up to 2 MHz and images were integrated over many beam pulse cycles, the acquisition times ranging from tens to several hundreds of seconds. In fact, during the evaluation of Gen1 TRION, gate times no shorter than 8 ns could be realized, whereas more recent gate pulsers have been able to define gate widths as short as 4 ns at a full repetition rate of 2 MHz. A cooled CCD camera views the image created at the phosphor of the gated intensifier. Within the TOF window of typically 500 ns, depending on the distance between neutron-source and detector and the delay and width of the gate pulse relative to the beam pulse trigger, the detector integrates all neutrons of a certain energy window into an image. By varying the delay time of the gate pulse, transmission images at any desirable neutron energy can be taken. Of course, to obtain images at different energies, successive runs at different gate delay times are required with Gen1 TRION, since only a single frame is available at a time. The experimental results presented in part III are all based on this single-frame detector. The next generations of TRION, described below, are presently being evaluated (Gen2) or in the process of being assembled (Gen3).

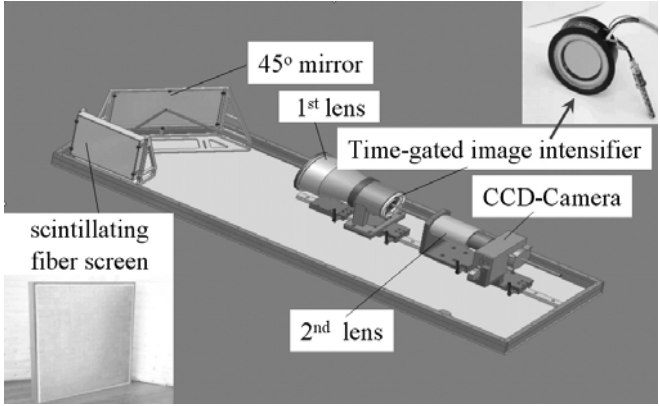

Fig. 2. Engineering drawing of single-energy TRION detector

### C. Multiple-Frame TRION Concepts

Using a single TOF-frame at a time in a white neutron spectrum, TRION is, of course, not very economical. Out of the broad neutron spectrum only the neutrons in a narrow energy window are employed, the others delivering an excessive radiation dose to the sample, to no useful effect. Therefore, to reduce dose and increase scanning speed, the goal is to make use of all available neutrons, if possible. To this end, we are developing 2 different systems for quasi-simultaneous neutron imaging in multiple pre-selectable energy (time) windows. In addition, a separate time window might also be allocated to imaging the gamma flash that precedes each neutron burst in the TOF spectrum obtained with a typical accelerator-driven pulsed neutron source.

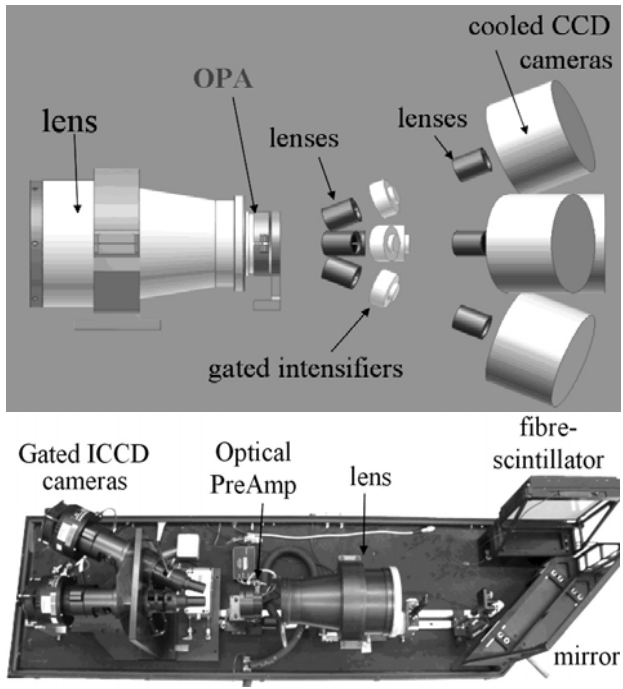

Fig. 3. Top: engineering drawing of a multiple-energy TRION detector, based on individual intensified cameras. Bottom: a photograph of a dual-energy system of this type.

To reduce cost and increase gating speed (the latter is important to achieve the required energy resolution at short neutron flight distances), small-area image intensifiers are preferential in front of each CCD camera. This runs counter to the requirement of efficient light detection, for which the demagnification of the optical image by the first lens should be as small as possible. Therefore, to achieve a high photodetection yield, a special Optical PreAmplifier (OPA) was proposed and jointly manufactured by two companies (Photek Ltd / UK [22] and ElMul / Israel [23]). In addition to detecting and intensifying the light image from the scintillation screen, this optical booster must preserve the fast timing property of the scintillator. This can only be achieved if the light decay time of the intensifier phosphor screen is also in the few-nanosecond range. Standard present-day image intensifier phosphors do not fulfill this criterion. ElMul Ltd. [23] offers such a fast phosphor for open multi-channel plate systems used in TOF electron microscopy. Fig. 4 shows the glow curves of two fast plastic scintillator screens and, for comparison, the one for the fast phosphor used in our OPA (dashed curve). The light decay constant of the phosphor screen is 2.4 ns, which is faster than the liquid and plastic scintillators that are candidates for the neutron converter screen to be used in this application (for the latter see also Lauck et al, these proceedings [24]). It should be mentioned that this phosphor also has a disadvantage which will require attention once the system is used at facilities with much higher neutron flux than in the presently available test facility at Physikalisch-Technische Bundesanstalt (PTB). The light yield of this phosphor is relatively low – only ca. 15 photons per 10 keV electron [24]. Therefore the OPA requires high electronic gain, which is a distinct drawback for high rate operation.

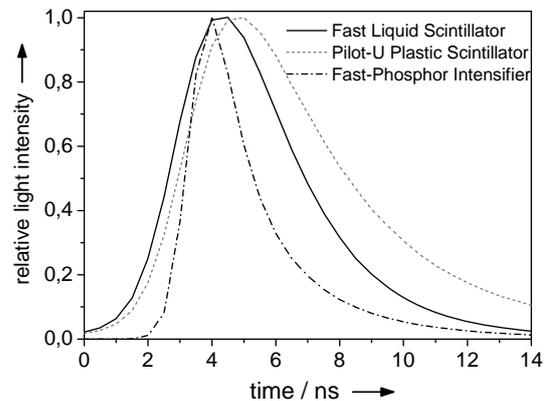

Fig. 4. Glow curve of two fast organic scintillators (solid and dotted curves) and, for comparison, that of the ultra-fast phosphor image intensifier, used here as an optical booster (dot-dashed curve).

Recently, a 4-camera system of this type was set up and tested at the neutron beam facility at PTB. Fig. 3 top shows an engineering drawing of this system and Fig. 3 bottom a photography of the previous dual-camera detector. Analysis of the results is in progress and instrumental improvements are presently being implemented to increase gain stability and gain monitoring of the intensifiers. Also, image processing needs to be given special attention, because the elemental reconstruction procedure requires pixel-by-pixel operations on the four different image projections. For this reason, image distortion effects due to the different projection angles call for appropriate corrective measures.

Another development geared for simultaneous imaging of up to 8 different energy frames is also briefly described here. However, since the critical electro-optical elements are now being produced, only the principle of operation will be presented here. It is based on a patented concept of a multi-frame high-speed optical camera system [25] and is now being optimized with the support of Invisible Vision, Ltd. [26] for anticipated use in energy-selective neutron radiography.

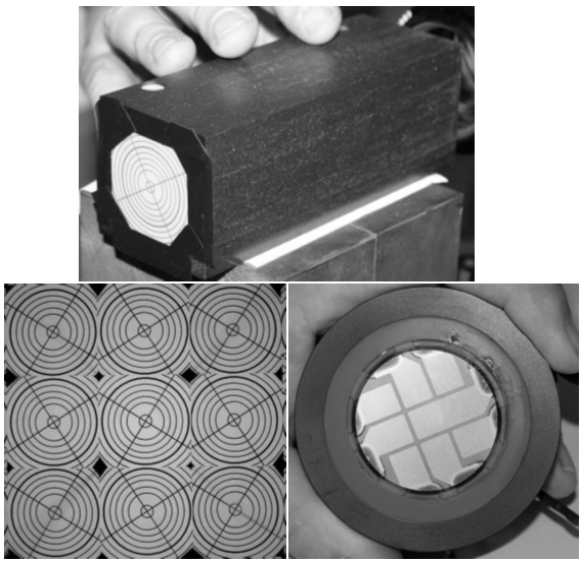

Fig. 5. Key components of the 8-frame camera system, consisting of a kaleidoscope-like image splitter (top) and an 8-fold segmented, independently gateable image intensifier (bottom right). The segmented image behind the image splitter is also shown (bottom left).

In this camera the image of the OPA (simulated by the bull's eye in Fig. 5 top), is viewed through a kaleidoscope-like image splitter, which produces an array of nine almost identical images of the same object (see Fig. 5, lower left). These nine images are projected by a relay lens to an independently gateable 8-fold segmented intensifier (see Fig. 5, lower right), whereby each segment views the scintillator screen and acquires an image for an independently selectable TOF region of the neutron/gamma burst.

To date we have only studied and optimized the individual components (image splitter, coupling lenses, fast UV-sensing intensifier, gating electronics, CCD camera) of such a system. Unfortunately, the available segmented intensifier of a decommissioned ULTRA8 high-speed camera is not well matched to the light of the optical booster, so that the production of an optimized segmented intensifier is being pursued by a commercial image intensifier manufacturer.

### III. EXPERIMENTAL RESULTS AND DISCUSSION

In this section we present results obtained with Gen1 TRION, which was extensively studied at the PTB neutron beam facility. Neutrons were produced by a 12 MeV deuterium beam impinging on a 3 mm thick Be target. The beam current was 2.5 μA (maximum current), the pulse width ca. 1.7 ns, the pulse repetition rate optionally 1 or 2 MHz and the TOF path ca. 12 m. The maximal neutron flux at 12 m distance is ca $2.5 \cdot 10^4$ cm$^{-2}$s$^{-1}$. The useful part of the neutron energy spectrum ranges from ca. 1 MeV up to 10 MeV (lower energy neutrons do not produce a measurable signal in the detector). More details about the PTB neutron facility and the neutron field employed here can be found in [27-29]

The position resolution was measured using a thick stainless steel mask which comprises slits of various width, spacing and thickness. Fig. 6 (top left) shows a photograph of the mask and the corresponding neutron transmission image (top right), integrated over all available energies between 1 and 10 MeV. Fig. 6, bottom, shows the measured Contrast Transfer Function (CTF), assuming 100 % CTF at a spatial frequency of $f_s$ = 0.05 line pairs per mm (lp/mm).

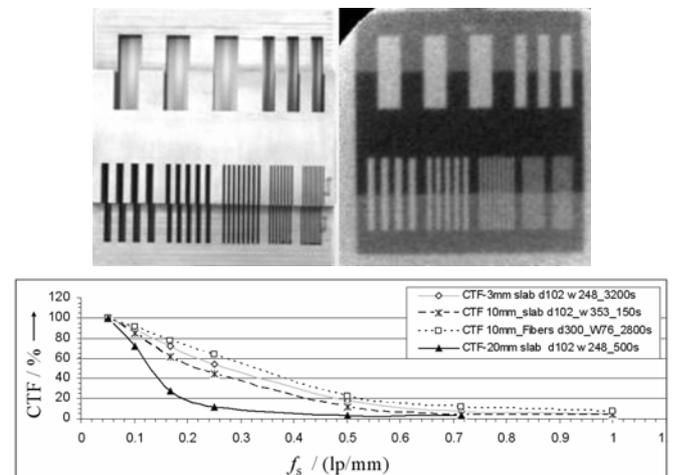

Fig. 6. Photograph (top left) and neutron transmission Image (top right) of a mask employed in measuring the Contrast Transfer Function (CTF). Bottom image shows the CTF as function of the spatial frequency $f_s$

Although the method is intended for automated detection of explosives, visual inspection of luggage and cargo contents is also important. As an example of the performance of the neutron imaging system and the potential for neutron interrogation, Fig. 7 shows a neutron transmission image of an ensemble of "suspicious" and potentially dangerous objects: small vials of acetone and water (simulating TATP), a toy pistol made of plastic, batteries and a mouth-piece of a brass musical instrument. Obviously the image resolution is as good as in comparable γ-ray images. But, unlike an X- or γ-ray image, which would only exhibit strong contrast for high-Z materials (such as the brass mouth-piece), high contrast for low-Z materials is also obtained. In another measurement, we have also demonstrated the detectability of structures composed of low-Z materials behind thick lead-shielding (see p.88 in [20]).

The key advantage of an automated explosives detection system (EDS) based on PFNTS is its capability for generating multi-elemental spatial distributions within an inspected object. This, in turn, requires acquisition of several (typically 6-8) images at specific neutron energies. Although Gen1

TRION is only able to produce these images sequentially, the simultaneous data acquisition of next-generation detectors can be experimentally simulated with the existing hardware, by accumulating images for different TOF-gate timings at the image intensifier.

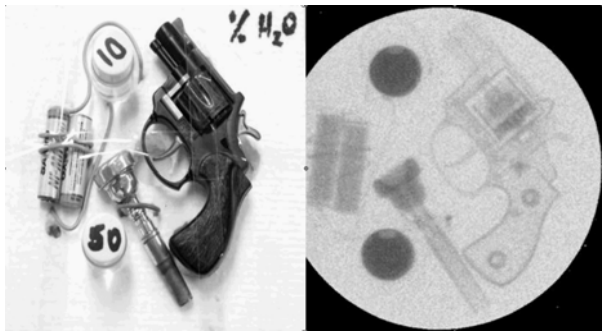

Fig. 7. Photograph and broad-energy neutron transmission image of a sample of various objects, including vials of acetone and water mixtures (that simulate TATP), a toy pistol made of plastic, batteries and a mouth-piece of a brass instrument.

Fig. 8 shows an object, consisting of melamine (which simulates here a commercial, nitrogen-rich explosive), several carbon rods and a steel wrench, that was consecutively imaged at various neutron energies. The top row of Fig 8 shows the neutron transmission images, marked with the corresponding TOF windows. At first glance they all look identical; however, with appropriate image arithmetic the net C and N distribution of the objects can be derived (Fig 8, lower right). Also, as expected, the steel wrench disappears completely in this process. Unfortunately, these images require very good event statistics in each of the individual TOF images. With the comparatively low-intensity accelerator beam and the technique of taking images sequentially, these elemental maps are hampered by poor neutron statistics.

## IV. SUMMARY

In this paper we have described TRION, a new integrative fast-neutron imaging method with energy selection using neutron-TOF. This method is well-suited for Pulsed Fast-Neutron Transmission Spectroscopy (PFNTS), a promising technique for an automated explosives detection system. In terms of its spatial resolution and event-rate handling capability, TRION is unique and superior to existing counting systems. We have presented experimental results obtained with the first generation detector, a single time-frame system. Multiple-energy selective imaging in a pulsed, broad-energy neutron beam was experimentally simulated and element-specific mapping has been demonstrated. Indeed, "conventional" visual inspection is feasible with TRION, due to its high spatial resolution and high contrast for low-Z materials, even if these are shielded by high-Z substances.

We also present two concepts for upgrading the single-frame TRION to a system capable of acquiring quasi-simultaneously 4 or even 8 frames. A crucial component of these multi-frame systems is a state-of-the-art optical preamplifier, that enables high neutron detection efficiency and fast gating capability. Whereas an intermediate system, employing 4 individual intensified cameras, is already in evaluation, a new version, comprising a very economical image splitter design, is also being set up. It will be capable of acquiring up to eight frames simultaneously on a single, specialized image intensifier.

ACKNOWLEDGEMENT

We thank M. Riches from Invisible Vision Ltd, A. Schon and E. Cheifetz from El-Mul Technologies and I. Cox from Photek LTD for their support and fruitful discussions.
This work was supported by the U.S. Transportation Security Administration (TSA).

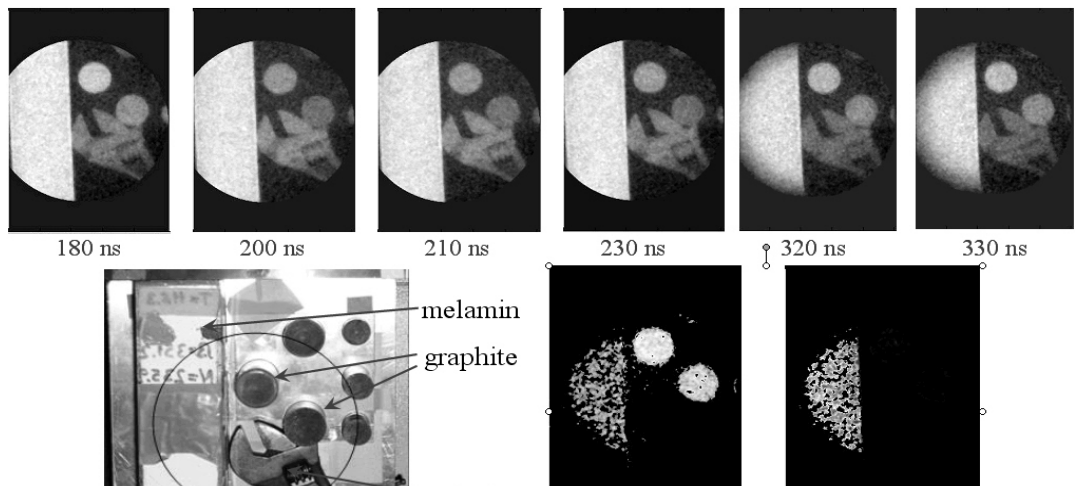

Fig. 8. Example of resonance imaging using TOF for energy selection. The bottom left image shows a photo of the sample. The circle marks the region which was imaged by a neutron detector (ca. 10 cm diameter). The top row of Fig. 8 shows 6 neutron images taken at different energies (TOF windows). From these images the net carbon and nitrogen distributions in the sample are derived (see the two images at bottom right).